\documentclass{IOS-Book-Article}

\usepackage{mathptmx}
\usepackage{graphicx}

\begin{document}
\begin{frontmatter}              

\title{Porting Large HPC Applications to\\
GPU Clusters: The Codes GENE and VERTEX}
\runningtitle{Porting Large HPC Application Codes to GPU Clusters}

\author[A]{\fnms{Tilman} \snm{Dannert}%
},
\author[A]{\fnms{Andreas} \snm{Marek}}
and
\author[A]{\fnms{Markus} \snm{Rampp}\thanks{Corresponding Author: Markus Rampp; E-mail:
markus.rampp@rzg.mpg.de.}}

\runningauthor{T. Dannert et al.}
\address[A]{Computing Centre (RZG) of the Max-Planck-Society\\and the
Max-Planck-Institute for Plasma Physics\\ Boltzmannstrasse 2, 85748 Garching, Germany.}

\begin{abstract}
We have developed GPU versions for two major high-performance-computing (HPC)
applications originating from two different scientific domains. 
GENE \cite{gene00,gene11} is a plasma microturbulence code which is employed
for simulations of nuclear fusion plasmas. 
VERTEX \cite{vertex02,vertex06,vertex13} is a neutrino-radiation 
hydrodynamics code for "first principles"-simulations of core-collapse 
supernova explosions \cite{snreview1,snreview2,snreview3}.
The codes are considered state of the art in their respective
scientific domains, both concerning their scientific scope and
functionality as well as the achievable compute performance, in particular 
parallel scalability on all relevant HPC platforms.
GENE and VERTEX were ported by us to HPC cluster architectures with 
two NVidia \emph{Kepler} GPUs mounted in each node in addition to two 
Intel Xeon CPUs of the \emph{Sandy Bridge} family.
On such platforms we achieve up to twofold gains in the overall
application performance in the sense of a reduction of the time to solution 
for a given setup with respect to a pure CPU cluster.    
The paper describes our basic porting strategies and benchmarking methodology,
and details the main algorithmic and technical challenges we faced on the 
new, heterogeneous architecture. 
\end{abstract}

\end{frontmatter}

\thispagestyle{empty}
\pagestyle{empty}

\section*{Introduction}
With GPU hardware and the corresponding software environments becoming mature,
compute clusters with GPU-accelerated nodes establish as a new, powerful
platform for high-performance computing (HPC). Mainly 
motivated by the expected boost for application performance
(i.e.\ reducing "time to solution") and also by energy-efficiency considerations (i.e.\ reducing "energy to solution"), major research
organizations and providers of HPC resources have already deployed an 
appreciable amount of GPU-accelerated resources worldwide
\cite{top500}. Moreover, GPU-like architectures are expected to play a 
major role in the upcoming exascale era \cite{iesp11}.

It is well known, however, in the community, that the new hardware architecture together
with the apparently disruptive programming models pose substantial
challenges to scientific application developers (e.g.\ \cite{iesp11}). While selected 
algorithms and applications have in fact been demonstrated to keep up with the shiny 
performance promises of GPUs, in some cases even at the very large
scale (e.g.\ \cite{shimokawabe11}), it remains to be seen whether a broader
class of scientific applications can take advantage of
GPU-accelerated systems with reasonable programming effort and in a sustainable way. Often inappropriately
termed "legacy applications" in this context, leading scientific HPC codes
are typically being actively developed, comprise many tens or hundreds
of thousands of lines
of code achieved by a team effort of many dozens of person years,
they provide state-of-the-art functionality, as well as high
optimization, parallel scalability and portability. The codes
GENE\footnote{GENE is developed by the group of F.\ Jenko (Max-Planck-Institute for Plasma
  Physics)} and VERTEX\footnote{VERTEX is developed by the group of H.-Th.\ Janka
  (Max-Planck-Institute for Astrophysics)}, which have been 
developed in the Max-Planck-Society with continuous support from 
its high-performance Computing Centre (RZG) may serve as prototypical examples 
in this respect.
But as a matter of
fact, such highly tuned codes are often reaching the limits of (strong) 
scalability. For example, due to increasing inter-node 
communication times, or, as in the case of VERTEX, due to the lack of
conventionally exploitable parallelism in the code structure, 
the time to solution for a given setup can no more
be efficiently reduced by utilizing more
CPU resources. Thus, a significantly increased node performance
due to accelerators appears as a promising route towards
further boosting application performances at scale.

\section{Methods}

Although both, GENE and VERTEX are written in FORTRAN (with MPI and
hybrid MPI/OpenMP parallelization, respectively) we decided to adopt the C-based CUDA
programming model because it is the performance reference 
for NVidia GPUs.
While we found the commercial CUDA-FORTRAN language to deliver
competitive performance on the GPU, the employed PGI compiler falls 
behind the Intel compiler (which is our reference for the
CPU) on the remaining CPU parts which marginalizes the overall
application speedups of the heterogeneous code. 
For the same reason we did not yet make productive use of the OpenACC 
programming model \cite{openacc}. 

The performance baseline for all comparisons is defined
by a highly optimized, parallel CPU implementation of the respective 
algorithms.
Rather than quoting single-core speedups (which, in our opinion is 
hardly meaningful in most cases) our comparisons are always based on 
the same number of GPU cards and \emph{multicore} CPUs
("sockets"). Specifically, we compare the run time obtained on a certain number
of nodes, each equipped with two Intel Xeon E5-2670 8-core CPUs and two 
NVidia K20X GPUs, with the run time
measured with the original, parallel CPU code on the same number of nodes
(without GPUs). 

\section{The GENE code}\label{sec:gene}

GENE \cite{gene00,gene11} is a massively parallel code for the simulation of plasma
turbulence in fusion devices. The code solves the time-dependent, five-dimensional
Vlasov-Maxwell system of equations on a fixed phase-space
grid. Depending on the physical problem, typical
GENE simulations take between a few days and many weeks, using
thousands of cores on x86\_64-based HPC systems. GENE is open-source 
\cite{genewww} and has a world-wide user base.

\subsection{Code structure}
\label{sec:gene_code_structure}

The GENE algorithm employs coordinates aligned to the magnetic field
lines in a fusion device like a tokamak.
In this paper we use the so called \textit{x-global} version, where
all physical quantities are handled in a spectral representation with
respect to the $y$ coordinate, which is the second of the three   
space dimensions $x$ (radial), $y$ (binormal) and $z$ (along the field
line). The remaining phase-space coordinates are (in this order) the
velocity along the field line $v_\|$ and the magnetic moment
$\mu$ (see \cite{gene11} for details). Although GENE is able to handle
any number of ion species and the electrons in the framework of
gyrokinetics, we use for this paper only a single ion species, neutralized
by electrons. For all performance comparisons a problem setup with a
number of $N_x=64, N_y=16, N_z=24, N_{v_\|}=96, N_\mu=16$ grid points
is used. 

The starting point of this work was a profiling of the GENE
code (SVN revision 3440), with the times given in
Table~\ref{tab:profiling_SNB}. 
\begin{table}[hb]
  \begin{tabular}{p{0.45\linewidth}p{0.45\linewidth}}
    \begin{minipage}[t]{.45\linewidth}
      \makebox[0cm]{}\\
      \begin{tabular}{l|ll}
        region & 1 SNB & 2 SNB\\
        \hline
        time loop & 13.5s & 6.9s\\
        field solver & * 2.0s & * 1.1s\\
        rhs computation & * 8.9s & * 4.4s\\
        nonlinearity & ** 5.4s & ** 2.7s
      \end{tabular}
    \end{minipage} 
    &
    \begin{minipage}[t]{.45\linewidth}
      \makebox[0cm]{}\\
      \begin{tabular}{l|rr}
        & 1 SNB & 2 SNB\\
        \hline
        CPU only & 4.4s & 2.1s\\
        Fermi M2090 & 7.4s & 3.5s\\
        Kepler K20X & 4.1s & 2.3s
      \end{tabular}
    \end{minipage}
    \\
    {\footnotesize \refstepcounter{table}\textbf{Table \thetable:} Profiling of GENE rev. 3440 on
      Sandy Bridge (SNB) sockets. The symbol * indicates the nesting level of the
      performance region.    \label{tab:profiling_SNB}} 
    &
    {\footnotesize \refstepcounter{table}\textbf{Table \thetable:} Performance of the
      nonlinearity with GPU acceleration.\label{tab:perf_comp}} 
  \end{tabular}
\end{table}

The times of Table~\ref{tab:profiling_SNB} show that the computation
of all terms of the right-hand side (rhs) of the Vlasov equation consumes
more than 60\% of the computing time in GENE, with the computation of the
(quadratic) nonlinearity dominating the other terms.

The computation of the nonlinearity follows the usual approach
\cite{canuto06} to avoid the computationally expensive convolutions in
spectral space by 
multiplication of the two fields corresponding to the quadratic term
in real space after Fourier transforming them. Specifically,
the fields are preprocessed (transposition, extension in $y$ direction
for dealiasing according to the $3/2$-rule) in a first step, and a
fast Fourier transform to real space is applied. 
After multiplication the result is transformed back to the spectral
representation, the additional modes are deleted, the array is
transposed and multiplied with a prefactor, and is finally added
to the right-hand side vector.

A high level of data parallelism ($\mathcal{O}(10^4)$) can be
realized, as the algorithm for the nonlinearity depends on the
remaining phase-space coordinates ($z$, $v_\|$ and $\mu$ and the
number of ion species) in a parametric way. This thread concurrency,
however, competes with regular MPI parallelism on the CPU: A larger
number of MPI ranks leaves less thread parallelism for the GPU and
vice versa.

\subsection{Algorithmic details and GPU implementation}
\label{sec:gene_porting}

The nonlinear term needs as input four arrays, the $x$ and $y$ component of the
$E\times B$-velocity and the derivatives of the distribution function
with respect to $x$ and $y$. Additionally it needs the already
computed right-hand side, to which the result is added after
multiplication with a prefactor. 

To minimize the transfer costs, we split each input array
into (usually four) contiguous chunks of $xy$-planes, which are
transferred and computed in two asynchronous CUDA streams. 
Additional asynchronicity comes from the fact that all MPI
tasks assigned to one CPU socket share one GPU. This also accounts for an overlap
of different kernels and transfer and helps to fully utilize the GPU card.  

The different parts which are described in
Sect.~\ref{sec:gene_code_structure} are implemented as a kernel
each (Tab.~\ref{tab:kernels_used}).
\begin{table}[!ht]
  \centering
  \begin{tabular}{|lp{0.6\linewidth}|}
    \hline
    \texttt{transpose}& transposes in $x$-$y$ dimension\\
    \texttt{extend\_for\_dealiasing} & add 50\% of zero modes which
    are filled by the multiplication\\
    \texttt{shrink\_for\_dealiasing}& after multiplication, remove the
    added modes to get the dealiased solution\\
    \texttt{compute\_nonlinearity}& multiplication in real space of
    the two arrays according to the structure of the nonlinearity\\
    \texttt{multiply\_with\_prefactor}& multiply the result with a
    prefactor and add to the original right-hand-side\\
    \texttt{CUFFT} & use this library for the Fourier transformations\\
    \hline
  \end{tabular}
  \\
  \caption{Kernels used for the GENE nonlinearity}
  \label{tab:kernels_used}
\end{table}
From profiling with the NVidia visual profiler, we find that all of
our kernels have a good coalescing memory access pattern and show a
high utilization of the GPU (occupancy $\geq$ 65\%). We also find that
more than half of the GPU 
run time is used by the Fourier transforms from the \texttt{CUFFT}
library (53\%), followed by \texttt{transpose} (17.9\%), 
\texttt{extend\_for\_dealiasing} (11.7\%), and
\texttt{compute\_nonlinearity} (8.8\%). 

\subsection{Performance results}
\label{sec:perf_results_gene}

The performance results for a Sandy Bridge CPU (8 cores) with
Fermi or Kepler GPU is shown in Table~\ref{tab:perf_comp}.
So far, we achieve only a very moderate acceleration
of the code by using a GPU. To get a deeper insight, we use the roofline
performance model \cite{roofline09}. We define as performance metric
for the nonlinearity of GENE the number of $xy$-planes computed per
second (unit: $\mathrm{P_c}/\mathrm{s}$) and as metric for the
bandwidth, we use the number of $xy$-planes transferred from host to
device via the PCI express bus (unit: $\mathrm{P_t}/\mathrm{s}$).  

Having defined the target metrics, it is necessary for the model to
assess a peak value (or ceiling) for these two metrices. The ceiling
for the bandwidth can be computed from the measured values of the
bandwidth of the PCIe bus, which is $5.7\,\mathrm{GB/s}$ (for PCIe
2.0). Therefore we get in our units (using $S_c=16\mathrm{B}$, the size of a
double-complex representation) 
\begin{displaymath}
  \mathrm{BW}_0=\frac{5.7\,\mathrm{GB/s}}{N_xN_y\cdot S_c}=374\,\mathrm{kP_t/s}
\end{displaymath}

\begin{figure}[!ht]
  \centering
  \begin{minipage}[t]{3.0in}
    \vspace{0pt}
    \includegraphics{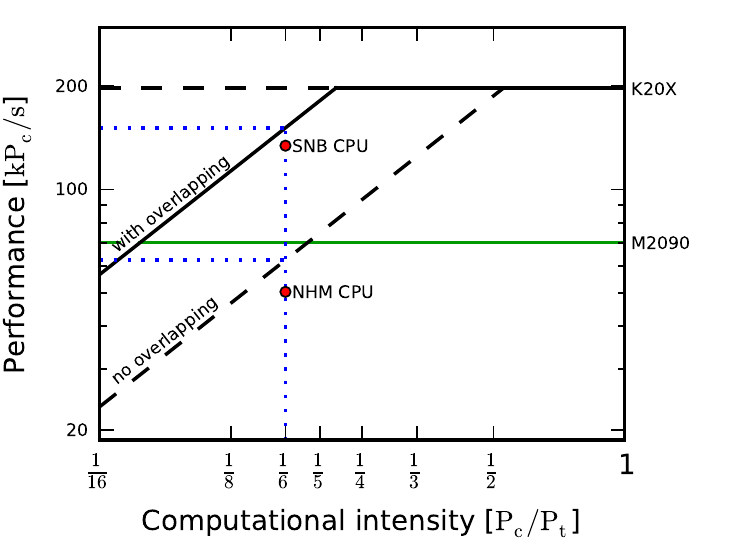}
  \end{minipage}\hfill
  \begin{minipage}[t]{1.85in}
    \caption{Roofline diagram for Kepler K20X
      (solid black) and Fermi M2090 (solid green). Bandwidth ceiling for
      synchronous transfer (dashed) and asynchronous transfer
      (solid). The nonlinearity is indicated by the dotted blue vertical
      line. Achieved CPU performance for Sandy Bridge (SNB, $134\,\mathrm{kP_c/s}$) and Nehalem
      (NHM, $50\,\mathrm{kP_c/s}$) is shown as red points for comparison.}
    \label{fig:roofline_basic}
  \end{minipage}
\end{figure}

The peak value for the computing performance is determined by measuring the
number of computed $xy$-pages, assuming the data is already present on the
GPU. Hence, after transferring all data to the device, synchronizing
the GPU with all MPI ranks, we time only the computation of the
$xy$-planes on the GPU. This number depends on the quality of the
kernels and not on the transfer and hence can define a ceiling for the computing
performance.
\begin{table}[!ht]
  \begin{minipage}[t]{0.5\linewidth}
    \makebox[0cm]{}\\
    \begin{tabular}{lrrr}
      & 8 MPI & 4 MPI & 2 MPI\\
      \hline
      time on K20X & 4.6s & 3.1s & 3.0s\\
      performance [$\mathrm{kP_c/s}$] & 128 & 190 & 197\\
      \hline
      time on M2090 & 8.4s & 11.0s & 11.3s\\
      performance [$\mathrm{kP_c/s}$] & 70 & 54 & 52
    \end{tabular}
  \end{minipage}\hfill
  \begin{minipage}[t]{0.45\linewidth}
    \makebox[0cm]{}\\
    {\footnotesize \refstepcounter{table}\textbf{Table \thetable:}
      Timing of the computation of 589824 $xy$-planes on a Kepler 
      K20X and a Fermi M2090 excluding data transfer. The results give the
      computing performance ceiling for the roofline model. 
      \label{tab:timing_peak}}
  \end{minipage}
\end{table}
Using these numbers (cf. Tab.~\ref{tab:timing_peak}) in our modified
roofline model, we obtain the solid horizontal line and the dashed line
labelled ``no overlapping'' in Fig.~\ref{fig:roofline_basic}. 

In a next step we switch on the asynchronous CUDA streams to hide the
transfer and measure the achieved total performance. The transfer time
decreases significantly, hence doubling the transfer bandwidth which gives a much
higher ceiling in the roofline graph (the solid, inclined bandwidth line). 
The computational intensity of the algorithm is $1/6$ (indicated by
the dotted blue vertical line in the figure), as we need 6 
transferred $xy$ planes (5 input, 1 output) to compute one result
$xy$-plane . 

For the K20X, the algorithm is still in the bandwidth-limited region,
hence we cannot exploit the full computational power of the GPU, whereas
for the older model (M2090) the limitation comes from the computational performance
of the GPU. In the latter case, faster kernels would help to get a
better performance, but for obvious reasons, we do not put effort
in optimizing for obsolete hardware. 
This roofline model shows clearly that when using the PCIexpress
bus, gen. 2.0, the performance of the nonlinearity on a Kepler K20X
card is bound by the data transfer between host and device. Hence, a
further optimization of the kernels will not help to improve the 
overall performance.  
This analysis also shows a way for further improvements. One can
try to overcome the bandwidth limit by moving more computation
to the GPU, while keeping the amount of transfer 
constant, or one uses the next generation 3 of the PCI bus, which
nearly doubles the bandwidth. In both cases, the algorithm will become compute-bound and
reaches the compute ceiling. Note that contrary to the
original roofline model \cite{roofline09}, this ceiling is not defined
``top-down'' from the nominal peak performance of the hardware, but is given by the
actual kernel performance. Using such an application-specific metric
facilitates the determination of the computational intensity in our case.
It is justified, provided that the analysis is confined to the bandwidth-limited
region of the roofline diagram.

\section{The VERTEX code}
VERTEX \cite{vertex02,vertex06} is a massively parallel,
multi-dimensional neutrino-radiation hydrodynamics code for simulating 
core-collapse supernova explosions \cite{snreview1,snreview2,snreview3}. Typical model runs 
require between a few months on 256 cores (for two-dimensional, axisymmetric
simulations) and up to 64\,000 cores (for the latest generation of
three-dimensional models) on HPC systems based on x86\_64 processors.

\subsection{Code structure}
\begin{figure*}[!t]
\begin{tabular}{cc}
 \resizebox{0.80\linewidth}{!}{\includegraphics{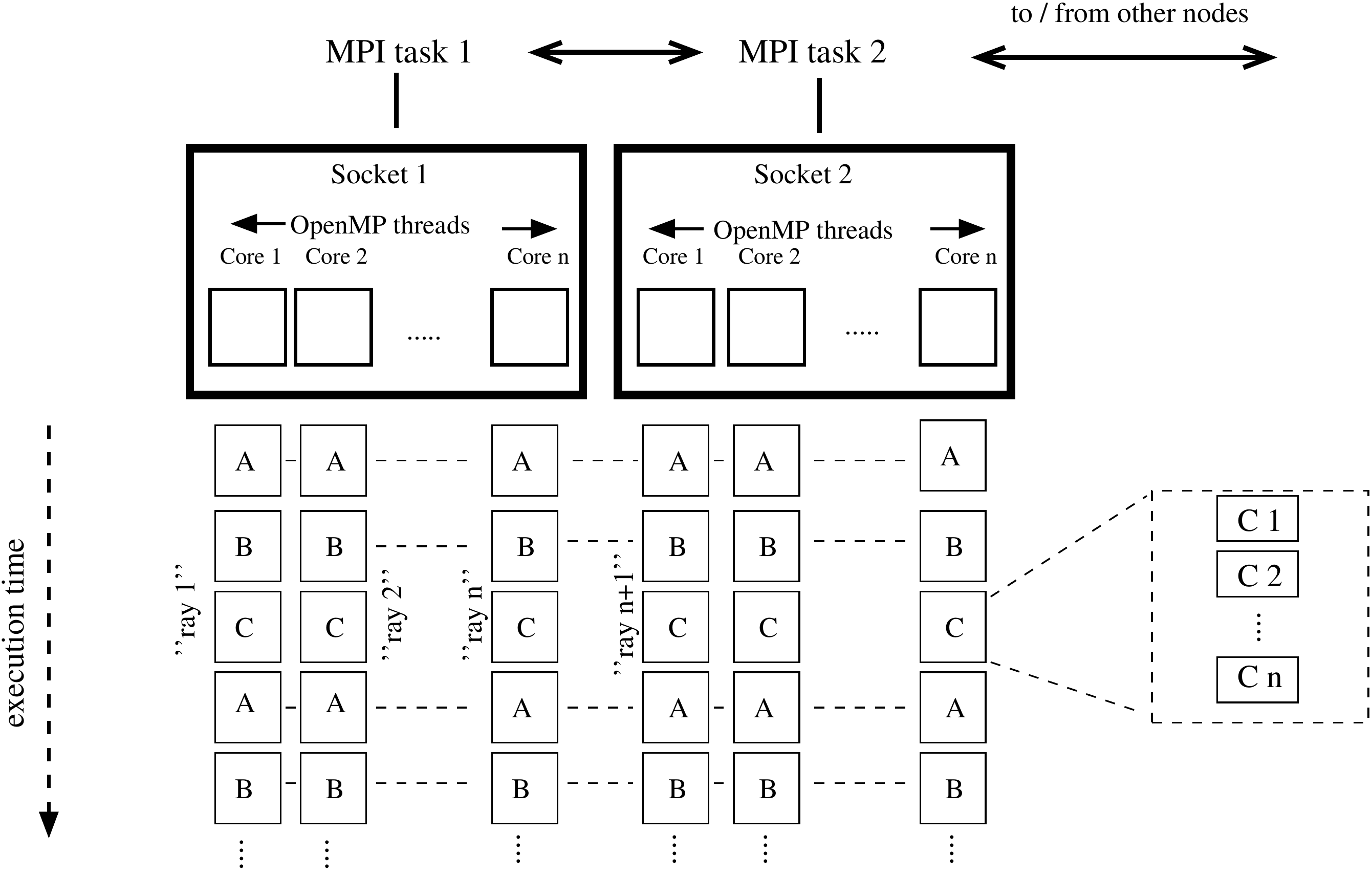}}
 & \\
  \bfseries{(a)} & \\
 \resizebox{0.55\linewidth}{!}{\includegraphics{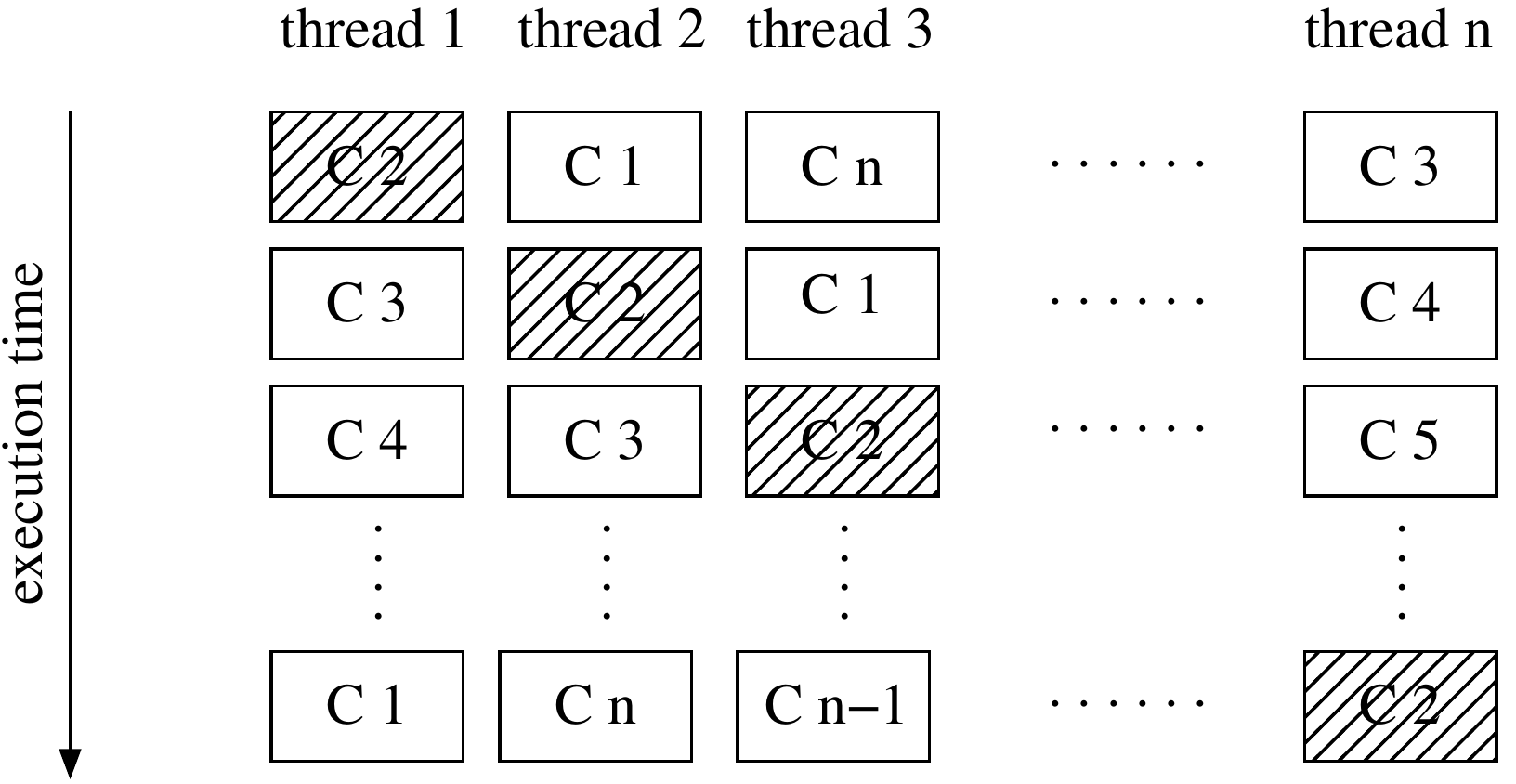}}
 & \\
 \bfseries{(b)} & \\
 \end{tabular}
 \caption{{\bfseries{(a)}:} Sketch of the parallelization approach of VERTEX and the mapping to nodes and cores. Typically, one MPI 
 task is mapped onto one socket. Within each socket OpenMP threads are
 pinned to the available cores. The figure also shows the execution flow of the code: each OpenMP thread executes the same sequence of steps during a time-step (indicated
 by the sequence \texttt{'A'},\texttt{'B'}, \texttt{'C'}). The blow up of step \texttt{'C'} 
 indicates the division into several sub-steps. Note, that part
 \texttt{'C'} and its substeps, which are the relevant parts for GPU
 acceleration, are not affected by any MPI synchronization (indicated by the
 dashed horizontal lines), and are completely independent of each other. {\bfseries{(b)}}: Sketch of
 the scheduling of concurrent work on one GPU and one CPU
 socket. White boxes indicate computations by one OpenMP thread, shaded
 boxes indicate the parts which are offload to the GPU.}\label{vertex:fig1}
\end{figure*}

VERTEX employs a spherical grid and a hybrid MPI$/$OpenMP
parallelization, based on a standard domain decomposition. Each MPI
domain is further divided into angular 'rays' (see Fig.~\ref{vertex:fig1}a) of
computational work by virtue of a coarse-grained OpenMP parallelization.
Figure~\ref{vertex:fig1}a schematically shows the division of the computational work into MPI tasks and 
OpenMP threads ('rays'), together with the mapping on the
hardware. The figure also sketches the execution flow of the
application, with different threads working independent of each other
during a time-step.

\begin{table}
\begin{tabular}{ll}
\parbox{.465\linewidth}{
\centering
  \begin{tabular}{|lll|}
  \hline
  code part   & 1 socket & 128 sockets\\ \hline
  \texttt{total}       & 4.56 s           &     4.63 s      \\
  \texttt{hydro} \texttt{(A)}      & * 0.05 s         &     * 0.10 s    \\
  \texttt{transport} \texttt{(B)}   & * 4.51 s        &     * 4.53 s    \\
  \texttt{rates} \texttt{(C)}      & ** 2.16 s       &     ** 2.16 s   \\
  \texttt{rate kernel} \texttt{(C2)}& *** 1.96 s      &     *** 1.96 s  \\ \hline
\end{tabular}
 } &
\parbox{.465\linewidth}{
\centering
\begin{tabular}{|ll|}
\hline
  code part & 1 socket \\ \hline
    \texttt{total}      &     2.62 s           \\
  \texttt{hydro} \texttt{(A)}     &    * 0.05 s       \\
  \texttt{transport} \texttt{(B)}  &   * 2.57 s        \\
  \texttt{rates} \texttt{(C)}      &   ** 0.19 s     \\
  \texttt{rate kernel}  \texttt{(C2)} &     *** 0.04 s      \\ \hline
\end{tabular}
} \\
 \bfseries(a) & \bfseries(b) \\

 \end{tabular}
 \tiny{\caption{\label{vertex:tab1} {\bfseries{(a)}}: Run time
     analysis for VERTEX in a weak-scaling setup on the CPU (Intel
     Sandy Bridge, 8 cores) using
 1 and 128 sockets, respectively. We show the total run time for
 one time step, together with a breakdown (inclusive times) of the main
 computational parts of the code. The labels \texttt{'A'},
 \texttt{'B'}, \texttt{'C'}, and \texttt{'C2'} allow a comparison with the execution flow sketched in Fig.~\ref{vertex:fig1}a.
 The symbol '*' indicates different nesting levels. The \texttt{'rate kernel'}
 consumes roughly 50 percent of the total run time. {\bfseries{(b)}}: Same as
 (a) but with GPU acceleration (one GPU per socket) enabled. 
 Note that only the part \texttt{'C2'} is computed on the GPU, but other sub-steps of \texttt{'C'} can be overlapped with the computation on the GPU. The prototype system with 
 two K20X GPUs allows to demonstrate weak scaling only up to two
 sockets (the times for 2 sockets are identical and not shown here). Given 
 the excellent weak scalability of VERTEX on CPU-based systems
 \cite{vertex_scaling, parco_vertex}
 our approach on the GPU allows to extrapolate the  measured speedups
 to large GPU-accelerated systems.}}
\end{table}
Table~\ref{vertex:tab1}a shows that the run time of VERTEX on the CPU is dominated by solving the radiative
transfer equations (item \texttt{'transport'}), and in particular for
computing neutrino absorption and emission rates (item \texttt{'rates'}). Fig.~\ref{vertex:fig1}a identifies the positions of the individual routines in the execution flow.
About 50 percent of the run time is spent in the computation of one particular interaction rate 
(named \texttt{'rate kernel'}, \texttt{'C2'}). The different interaction rates are often termed 
"local physics", which expresses the fact that the computations are to a high degree 
independent of each other and provide a data parallelism on the grid level. Different interaction processes (\texttt{'rates'}) can be computed independently of
each other, which implies additional, coarse-grained parallelism on the function level (see blowup in Fig.~\ref{vertex:fig1}a).

\subsection{Algorithmic details and GPU implementation}

In the following the algorithm for offloading the \texttt{'rate kernel'} to the
GPU is outlined. Due to its dominance in the code, high data parallelism
and arithmetic intensity the suitability for the GPU shall become immediately apparent.

As input for the computations a few one-dimensional arrays are needed, which represent the local 
thermodynamic conditions for which the interaction kernel is evaluated. All 
operations are performed on a five-dimensional grid representing discretized phase space. The size of this grid varies with the resolution, in a typical setup the total 
number of grid points is about $150\times10^6$. For the major part of
the kernel, computations on each grid zone can be done independently
of the others, which leads to a high degree of data parallelism (up to
$\mathcal{O}(10^5)$ threads). Only after all grid zones are
processed, a reduction (corresponding to a phase space integral) to a three 
dimensional grid is performed. This can be still done in
parallel, but with much less parallelism ($\mathcal{O}(10^3)$ threads). All computations are done
twice for subsets of different input data, accounting for two possible reaction channels.

The actual implementation of the part \texttt{'C2'} is straightforward: the data is copied asynchronously to the GPU and 
the five-fold nested loops of the CPU version are separated in kernel
calls with about 100\,000 threads. The kernels are scheduled in streams, in
order to allow the CUDA run time to overlap kernel executions corresponding to the
twofold computation of the processes. The problem is compute bound,
as data transfer is negligible (0.9 ms) compared to GPU
computations (40 ms) and at least 140 double-precision floating-point operations are executed per transferred byte.

For good performance results it turned out to be crucial to
use shared memory for the input data and to use as much registers as
possible on the device. After tuning our
CUDA code with the help of the NVidia profiler, we achieve an occupancy of 93\%
of the theoretical upper limit for the
most important kernels. However, we still encounter about 10\% of branch-divergence overhead and 25\% of global memory replay overhead. Work is
still ongoing to improve on the latter performance metrics.

As mentioned above, the different sub-steps \texttt{'C1'} to \texttt{'Cn'} (see Fig.~\ref{vertex:fig1}b) are 
independent of each other and can be computed in any order within one OpenMP thread, or 'ray'. In the original code, however, the order across different OpenMP threads
 is always the same, e.g. when a thread computes sub-step \texttt{'C1'}, also the other threads work on the same sub-step. An overlap of computations on the cores and the GPU was thus 
achieved by: a) individually shuffling the computations of the sub-steps \texttt{'C1'} to \texttt{'Cn'} on each 'ray', and b)
ensuring that the sub-step \texttt{'C2'} from each 'ray' to the GPU is offloaded in a queue (see Fig.~\ref{vertex:fig1}b).
In an ideal situation where all steps  \texttt{'C1'} to \texttt{'Cn'} take the same amount of execution time, work on the CPUs and the
GPU would be perfectly overlapped. In reality, a balancing of about 80\% could be reached.

\subsection{Performance results}
The rate kernel \texttt{'C2'} requires 2.16~s on one CPU thread (cf. Tab~\ref{vertex:tab1}) and scales almost perfectly with OpenMP. The same kernel can be
computed on the GPU in 0.04~s. Thus, with one GPU, speedups of 7 or 54 are
achieved when comparing with one CPU socket or a single core, respectively.
This demonstrates that a significant speedup was achieved with respect to a
Sandy Bridge CPU. As the coarse grained OpenMP parallelization of VERTEX (which
is crucial for achieving its excellent weak scalability) does not allow to
use the threaded rate kernel on the CPU, the acceleration factor of 54 applies
for production applications which effectively eliminates the rate kernel from
the computing time budget and in practice accounts for a twofold 
acceleration (corresponding to the original 50\% share of the rate kernel, cf. Tab~\ref{vertex:tab1}) of the 
entire application.

\section{Summary and Conclusions}

With the specific cases of GENE and VERTEX we have shown that
complex HPC applications can successfully be ported to 
heterogeneous CPU-GPU clusters. 
Besides writing fast GPU code, exploiting and balancing both 
the GPU \emph{and} the CPU resources of the heterogeneous compute nodes 
turned out to be an essential prerequisite for achieving good overall "speedups", which we
define as the ratio of the run time obtained on a number
of GPU-accelerated sockets and the run time measured with 
parallel code on the same number of CPU sockets. 

In the case of VERTEX we have demonstrated twofold speedups which hold 
for production applications on GPU-clusters with many hundreds of
nodes. In particular, the excellent weak scalability of VERTEX \cite{parco_vertex}
is not affected by the additional
acceleration due to GPUs. Threefold speedups appear in reach but would require at least 
additional porting of a linear solver for a block-tridiagonal system. 
Limitations in the software 
environment (lack of device-callable LAPACK functionality) have so far 
impeded a successful port of this part of the algorithm. Importantly, 
due to the specific code structure of VERTEX, such speedups would not 
have been possible with comparable
programming effort by simply using more CPU cores.

The performance of GENE is currently limited by the data transfer between
the host CPU and the GPU as we have shown by an elaborate performance-modeling 
analysis. After this bottleneck will have relaxed by upcoming 
hardware improvements (PCIe 3) further optimization efforts on the
GPU code will increase the overall speedups on a heterogeneous cluster.

The question whether the effort of several person-months,
which we have invested for each code, and which we consider typical
for such projects, is well justified cannot be answered straightforwardly. 
For complex, and "living" scientific HPC codes, for which GENE and
VERTEX can serve as prototypical examples, achieving up to threefold speedups in 
overall application performance appears very competitive 
\cite{acss12}. Also from the point of view of
hardware investment (buying GPUs instead of CPUs) and operational 
costs ("energy to solution") the migration of applications 
from pure CPU machines to GPU-accelerated clusters can be considered 
cost-effective if speedups of at least about two are achieved.
On the other hand, while very valuable for increasing simulation
throughput, twofold or threefold application speedups usually do not
enable qualitatively new science objectives. For this reason we
sometimes observe reluctance in the scientific community to invest significant 
human resources for achieving GPU-performance improvements in this
range. This is further exacerbated by legitimate concerns about sustainability,
maintainability and portability of GPU-kernel code. 
These are no serious issues for GENE and VERTEX, where the parts we
have ported to the GPU are not under heavy algorithmic development and
were also carefully encapsulated by us. 
In general, however, the \emph{need} for kernel programming, which is
considered as a pain by many, currently appears as the largest hurdle for a 
broader adoption of GPU programming in the scientific HPC community.
Moreover, it may turn out necessary to port significant parts of 
the application code to the GPU, e.g. in cases like GENE where the
data transfers become a limiting factor, or even to completely
reimplement the application.
 
These concerns could be mitigated by the establishment of a high-level, 
directive based programming model, based e.g.\ on the OpenACC standard \cite{openacc}
or a future revision of OpenMP \cite{openmpacc}, together with
appropriate compiler support. Also Intel's Xeon Phi many-core
coprocessor with its less disruptive programming model appears
very prospective in this respect. Despite serious efforts, however, we 
were not yet successful with GENE or VERTEX to achieve
performances on this platform which are competitive with the GPU. 
We attribute this mostly to a comparably lower maturity of
the Xeon Phi software stack and we expect improvements with
upcoming versions of the compiler and the OpenMP run time.

Most importantly, today's GPUs (and many-core coprocessors) might provide a
first glimpse on the architecture and the related programming
challenges of future HPC architectures of the exascale
era \cite{iesp11}. Applications need to be prepared \emph{in time} for
the massive SIMT and SIMD parallelism which is expected to become
prevalent in such systems.  
Even on contemporary multicore CPUs with comparably moderate thread-counts 
and SIMD width, the experience we have gained with porting GENE and
VERTEX has already led to appreciable performance improvements of the CPU codes.

\paragraph{Acknowledgments}
We thank F.\ Jenko and H.-Th.\ Janka for encouraging the development of 
GPU versions for GENE and VERTEX, respectively. 
NVidia Corp.\ and Intel Corp.\ are acknowledged for providing hardware samples and 
technical consulting.

\bibliographystyle{unsrt}

\end{document}